\documentstyle[11pt,aasms4,flushrt]{article}

\slugcomment{\it Appeared in Nature, 408, 560-562 (30 Nov 2000)} 
\lefthead{Stern et al.}
\righthead{Probable Redshift $z=6.68$ --- NOT.}

\def\ie{{i.e.,~}}
\def\eg{{e.g.,~}}
\def\etal{{et al.}}

\def\stis{STIS~123627+621755}

\def\deg{\ifmmode {^{\circ}}\else {$^\circ$}\fi}

\def\secper{\ifmmode \rlap.{^{s}}\else $\rlap{.}{^{s}} $\fi}

\def\kms{\ifmmode {\rm\,km\,s^{-1}}\else
    ${\rm\,km\,s^{-1}}$\fi}
\def\kmsMpc{\ifmmode {\rm\,km\,s^{-1}\,Mpc^{-1}}\else
    ${\rm\,km\,s^{-1}\,Mpc^{-1}}$\fi}
\def\ergAcm2{\ifmmode {\rm\,ergs\,cm^{-2}\,{\rm \AA}^{-1}}\else
    ${\rm\,ergs\,cm^{-2}\,\AA^{-1}}$\fi}
\def\ergcm2s{\ifmmode {\rm\,ergs\,cm^{-2}\,s^{-1}}\else
    ${\rm\,ergs\,cm^{-2}\,s^{-1}}$\fi}
\def\ergsHz{\ifmmode {\rm\,ergs\,s^{-1}\,Hz^{-1}}\else
    ${\rm\,ergs\,s^{-1}\,Hz^{-1}}$\fi}
\def\ergs{\ifmmode {\rm\,ergs\,s^{-1}}\else
    ${\rm\,ergs\,s^{-1}}$\fi}
\def\ergsA{\ifmmode {\rm\,ergs\,s^{-1}\,\AA^{-1}}\else
    ${\rm\,ergs\,s^{-1}\,\AA^{-1}}$\fi}
\def\WHz{\ifmmode {\rm\,W\,Hz^{-1}}\else
    ${\rm\,W\,Hz^{-1}}$\fi}
\def\WHzsr{\ifmmode {\rm\,W\,Hz^{-1}\,sr^{-1}}\else
    ${\rm\,W\,Hz^{-1}\,sr^{-1}}$\fi}
\def\ergscm2Hz{\ifmmode {\rm\,ergs\,cm^{-2}\,s^{-1}\,Hz^{-1}}\else
    ${\rm\,ergs\,cm^{-2}\,s^{-1}\,Hz^{-1}}$\fi}

\def\spose#1{\hbox to 0pt{#1\hss}}
\def\simlt{\mathrel{\spose{\lower 3pt\hbox{$\mathchar"218$}}
     \raise 2.0pt\hbox{$\mathchar"13C$}}}
\def\simgt{\mathrel{\spose{\lower 3pt\hbox{$\mathchar"218$}}
     \raise 2.0pt\hbox{$\mathchar"13E$}}}

\def\lyb{Ly$\beta$}

\def\lya{Ly$\alpha$}

\def\oii{[\ion{O}{2}] $\lambda$3727}

\def\plotfiddle#1#2#3#4#5#6#7{\centering \leavevmode
\vbox to#2{\rule{0pt}{#2}}
\includegraphics{#1}}

\begin{document}

\title{Evidence Against a Redshift $z > 6$ for the 
Galaxy \stis\altaffilmark{1}}

\author{Daniel Stern\altaffilmark{2},
Peter Eisenhardt\altaffilmark{2},
Hyron Spinrad\altaffilmark{3},
Steve Dawson\altaffilmark{3}, \\
Wil van~Breugel\altaffilmark{4}, 
Arjun Dey\altaffilmark{5},
Wim de~Vries\altaffilmark{4},
\& S.~A.~Stanford\altaffilmark{4,6}}

\altaffiltext{1}{Based on observations at the W.M. Keck Observatory,
which is operated as a scientific partnership among the University of
California, the California Institute of Technology, and the National
Aeronautics and Space Administration.  The Observatory was made possible
by the generous financial support of the W.M. Keck Foundation.}

\altaffiltext{2}{Jet Propulsion Laboratory, California
Institute of Technology, Mail Stop 169-327, Pasadena, CA 91109 USA}

\altaffiltext{3}{Department of Astronomy, University of California at
Berkeley, Berkeley, CA 94720 USA}

\altaffiltext{4}{Institute of Geophysics and Planetary Physics,
Lawrence Livermore National Laboratory, L-413, Livermore, CA 94550 USA}

\altaffiltext{5}{KPNO/NOAO, 850 N. Cherry Ave., P.O. Box 26732, Tucson,
AZ 85726 USA}

\altaffiltext{6}{Physics Department, University of California at Davis,
Davis, CA 95616 USA}

\bigskip
\bigskip
\bigskip
\bigskip
\begin{center}
{\it Appeared in Nature, 408, 560-562 (30 Nov 2000)}
\end{center}
\bigskip
\bigskip
\bigskip
\bigskip
\bigskip
\bigskip
\bigskip
\bigskip
\bigskip
\bigskip
\bigskip
\bigskip
\bigskip
\bigskip
\bigskip
\bigskip
\bigskip
\bigskip
\bigskip
\bigskip
\eject

%\begin{abstract}
%{\center NO ABSTRACT FOR NATURE LETTERS}
%\end{abstract}

{\bf 

The identification of galaxies at extreme distances provides our most
direct information on the earliest phases of galaxy formation.  The
distance implied by redshifts $z > 5$ makes this a challenging endeavor
for even the most luminous sources; interpretation of low
signal-to-noise ratio observations of faint galaxies is difficult and
occasional misidentifications will occur.  Here we report on \stis, a
galaxy with a suggested spectroscopic redshift of $z = 6.68$ [1] and
the most distant spectroscopically-identified object claimed.  For the
suggested redshift and reported spectral energy distribution, the
galaxy should be essentially invisible at $\lambda < 9300$~\AA\ because
the intervening intergalactic medium absorbs essentially all light
energetic enough to ionize neutral hydrogen ($\lambda < (1 + z) \times
912$ \AA; the redshifted Lyman limit).  The galaxy should be relatively
bright in the near-infrared with $f_\nu \simeq 1 \mu$Jy.  Here we
report a detection of the galaxy at 6700 \AA, below the Lyman limit for
$z = 6.68$.  We also report a non-detection at 1.2 $\mu$m, implying the
flux has dropped by a factor of three or more between rest
1216~\AA\ and rest 1560~\AA\ for $z = 6.68$.  Our observations are
inconsistent with the suggested extreme distance of \stis\ and
conservatively require $z < 6$.

}

During UT 1997 December 23-26, while the Wide Field Planetary Camera~2
obtained second epoch images of the Hubble Deep Field$^2$ (HDF), the
Space Telescope Imaging Spectrogragh (STIS) obtained 4.5~hr of
filterless imaging ($\lambda_c = 5850$ \AA; FWHM = 4410 \AA) and
13.5~hr of slitless spectroscopy of a $52\arcsec \times 52\arcsec$
field 4.7~arcmin NNW of the HDF.  The STIS field is sufficiently
separated from the HDF that it lies outside the initial HDF flanking
field imaging surveys; \ie no supporting photometry is provided by the
single orbit $I_{814}$ images of Williams \etal$^2$, the deep Hawaii
2.2~m $V,I,H+K$ images Barger \etal$^3$, or the deep $U_n,G, {\cal R}$
images of Steidel \etal$^4$.  Chen \etal$^1$ have analyzed the deep
STIS data to identify distant protogalaxies and report that one faint
source, \stis\ (``galaxy A'' in the discovery paper), has a {\em
probable} redshift of $z = 6.68 \pm 0.005$, which would make it the
most distant object identified spectroscopically.  The source is
marginally resolved and has $AB\rm{(clear)} = 27.7 \pm 0.1$.  The
redshift is based upon an emission line at $9337 \pm 6$ \AA\ with a
flux density of $2.6 \pm 0.5 \times 10^{-17} \ergcm2s$, identified as
\lya, and an extremely strong continuum break at $\approx 9300$ \AA,
identified as absorption from the \lya\ forest.

For $z = 6.68$, the flux detected in the filterless image must derive
almost exclusively from wavelengths longward of the \lya\ forest,
implying \stis\ has $f_\nu \approx 1 \mu$Jy (AB magnitude $\approx 24$)
longward of 9300 \AA, and should be essentially undetectable at optical
wavelengths shortward of 9300 \AA.  Assuming this source is a
star-forming galaxy as suggested by Chen \etal$^1$, the inferred 1700
\AA\ absolute magnitude is $M_{\rm AB}(1700 {\rm \AA}) - 5 \log h_{50}
= -22.3$ (assuming a flat spectrum source longward of \lya, $f_\nu
\propto \nu^0$, and $H_0 = 50~ h_{50}~ \kmsMpc$, $\Omega = 1$, $\Lambda
= 0$).  This is more luminous than all but a handful of the more than
700 galaxies at $z > 2$ identified by Steidel and collaborators$^5$
over a field-of-view more than a 1000 times larger.  If \stis\ is a
quasar at $z = 6.68$, the inferred blue luminosity is less atypical of
the population with $M_B \approx -22.9$ (assuming a standard quasar
optical spectral index of $f_\nu \propto \nu^{-0.5}$).  However, the
probability of identifying a faint ($M_B \leq -22.5$) quasar at $z \geq
6$ in the STIS field-of-view is $\approx 3 \times 10^{-4}$ (\eg see
Stern \etal$^6$).

These calculations show that \stis\ is potentially a very unusual and
interesting source, providing us with a uniquely luminous probe of the
very early Universe.  We therefore targeted it for deep studies with
the Keck telescopes (Fig.~1).  We do not detect \stis\ in our $J$-band
image, implying $J_{\rm AB} > 25.3$ (3$\sigma$ limit in a 2\farcs0
diameter aperture).  We report a 6$\sigma$ detection of \stis\ in our
$R$-band image:  $R_{\rm AB} = 26.8 \pm 0.1$ measured in a 0\farcs9
diameter aperture to avoid the neighboring galaxy to the east with
aperture corrections applied assuming the source is unresolved.
Between 1998 and 2000 we have also attempted slitmask spectroscopy of
\stis\ numerous times, never robustly detecting the source.  As an
example of one observation with good conditions and confident
astrometry, on UT 2000 April 06 we observed \stis\ for 2 hours with the
Low Resolution Imaging Spectrometer$^{8}$ and the 150 lines mm$^{-1}$
grating.  Our 3$\sigma$ limit to spatially (1\farcs0 FWHM) and
spectrally ($\sigma \approx 230 \kms$) unresolved line emission at 9350
\AA\ is $\approx 2 \times 10^{-17}$ \ergcm2s, marginally at odds with
the Chen \etal$^1$ results.  Note that the fringing and telluric OH
emission which afflict ground-based CCD observations at long wavelength
make this limit wavelength dependent.

Our results strongly imply \stis\ cannot be at $z > 6$.  In Fig.~2 we
present the photometric and spectrophotometric data for
\stis\ overplotted with a model star-forming galaxy at $z = 6.68$.  Our
1.2 $\mu$m non-detection is severely at odds with the spectral energy
distribution reported by Chen \etal$^1$, who predict a near-infrared
magnitude approximately twice as bright.  For the 9700 \AA\ flux
density reported from the slitless spectrum, our 1.2 $\mu$m
non-detection implies an extremely blue continuum slope, $\alpha_{\rm
UV} > 4.7$ (95\% confidence limit) where $f_\nu \propto
\nu^{\alpha_{\rm UV}}$.  This is significantly steeper than the
ultraviolet spectra of known star-forming galaxies$^9$:  longward of
\lya, the average spectral slope is $\alpha_{\rm UV} = -0.3$ and few
sources are bluer than $\alpha_{\rm UV} = 1.5$.  The slope implied by
our observations is also bluer than the Rayleigh-Jeans tail of a black
body spectrum ($f_\nu \propto \nu^2$), requiring either photometric
errors, variability, or a non-thermal origin to the flux.

Convolving the star-forming galaxy model (Fig.~2) with an $R$-band
filter transmission and detector response, we predict that a $z = 6.68$
galaxy with $f_\nu \approx 0.67 \mu$Jy longward of \lya\ should have
$R_{\rm AB} > 32$.  The $R$-band detection more than 100 times brighter
than the prediction implies that \stis\ {\em cannot} be at $z = 6.68$.
Furthermore, much of the STIS filterless imaging flux must derive from
wavelengths {\em shortward} of 9300 \AA, implying that the
spectrophotometric decrement at 9300 \AA\ must be significantly less
extreme than suggested.  This is consistent with our 1.2$\mu$m
non-detection and implies the 1$\mu$m spectral slope is not as extreme
as discussed above.

What is the likely redshift for \stis?  The marginally-resolved STIS
morphology, reported emission line, red $AB\rm{(clear)} - R$ color, and
blue ground-based color argue against a Galactic origin for the
source.  In particular, for $R-J < 1.5$, a stellar interpretation
conservatively requires a classification earlier than K5 (\eg Table
3.10 of Binney \& Merrifield$^{10}$), implying a distance $\simgt$
100~kpc for luminosity class V.  A faint Galactic white dwarf
identification is not ruled out by the ground-based color.  If
\stis\ is comparable to the lowest luminosity white dwarf known$^{11}$,
ESO~439$-$26 with $M_V \simeq 17.5$, the implied distance is $\approx$
1~kpc.  Most white dwarfs are 1.5$-$8 magnitudes more luminous$^{12}$.

The imaging results are consistent with a flat spectrum source of
$\approx 27$ magnitude.  If we assume the isolated emission line
detected at 9337 \AA\ is real, then \oii\ is the more likely
identification and \stis\ is similar in luminosity ($M_B \approx -17$)
to the Small Magellanic Cloud at $z = 1.51$ (see Fig.~2).  If so, the
inferred restframe \oii\ equivalent width is $W_{\rm [OII]} \approx 4$
\AA, typical compared to local samples of star-forming
galaxies$^{15}$.   The flux decrement at 9300~\AA\ is reproduced by
this model, albeit at a less dramatic amplitude, and is associated with
hydrogen Balmer absorption rather than hydrogen Lyman absorption.  The
blue $R - J$ color and failure of our extensive spectroscopy to detect
any features both support this interpretation.  However, the
$AB\rm{(clear)} - R$ color is $\approx 1$ magnitude too red and
suggests either variability or a spectral break in the range
$\lambda\lambda 3000 - 6000$\AA.

These results and other recent serendipitously-identified
sources$^{13,14}$ show that low luminosity, [\ion{O}{2}]-emitting
galaxies at $z \approx 1.5$ can be convincing doppelg\"angers of
high-redshift galaxies.  At $z \approx 3$ we know that the faint end of
the galaxy luminosity function is steep$^5$.  Low luminosity,
[\ion{O}{2}]-emitting galaxies  are likely to be a significant
contaminant to high-redshift \lya\ searches, suggesting that such
surveys should incorporate deep imaging bluewards of the emission line
to distinguish the populations.

\begin{center}
{\bf References}
\end{center}

1.  Chen, H.-W., Lanzetta, K.~M., \& Pascarelle, S.  Spectroscopic
identification of a galaxy at a probable redshift of $z = 6.68$.  {\it
Nature}, {\bf 398}, $586-588$ (1999).

2.  Williams, R.~E. {\it et~al.}  The Hubble Deep Field:  observations,
data reduction, and galaxy photometry.  {\it Astron. J.}, {\bf 112},
$1335-1389$ (1996).

3.  Barger, A., Cowie, L.~L., Trentham, N., Fulton, E., Hu, E.~M.,
Songaila, A., \& Hall, D.  Constraints on the early formation of field
elliptical galaxies.  {\it Astron. J.}, {\bf 117}, $102-110$ (1999).

4.  Steidel, C.~S., Giavalisco, M., Pettini, M., Dickinson, M., \&
Adelberger, K.~L.  Spectroscopic confirmation of a population of normal
star-forming galaxies at redshifts $z > 3$.  {\it Astrophys. J.}, {\bf
462}, L17$-$L21 (1996).

5.  Steidel, C.~S., Adelberger, K.~L., Giavalisco, M., Dickinson, M.,
\& Pettini, M.  Lyman-break galaxies at $z \simgt 4$ and the evolution
of the ultraviolet luminosity density at high redshift.  {\it
Astrophys.  J.}, {\bf 519}, $1-17$ (1999).

6.  Stern, D., Spinrad, H., Eisenhardt, P., Bunker, A.~J., Dawson, S.,
Stanford, S.~A., \& Elston, R.  Discovery of a color-selected quasar at
$z=5.50$.  {\it Astrophys. J.}, {\bf 533}, L75$-$L78 (2000a).

7.  Matthews, K. \& Soifer, B.~T.  in {\it Infrared Astronomy with
Arrays:  The Next Generation} (ed McLean, I.) 239$-$246 (Kluwer,
Dordrecht, 1994).

8.  Oke, J.~B.  The Keck low-resolution imaging spectrometer.  {\it
Publ. Astron. Soc. Pacific}, {\bf 107}, $375-385$ (1995).

9.  Adelberger, K.~L. \& Steidel, C.~C.  Constraints on dusty star
formation at high redshift from ultraviolet, far-infrared, and radio
surveys.  {\it Astrophys. J.}, in press, astro-ph/0001126 (2000).

10.  Binney, J. \& Merrifield, M.  {\it Galactic Astronomy} (Princeton 
Univ. Press, Princeton, 1998).

11.  Ruiz, M., Bergeron, P., Leggett, S., \& Anguita, C.  The extremely
low luminosity white dwarf ESO~439$-$26.  {\it Astrophys. J.}, {\bf
455}, L159$-$L162 (1995).

12.  Harris, H., Dahn, C., Vrba, F., Henden, A., Liebert, J., Schmidt,
G., \& Reid, I.  A very low luminosity, very cool, DC white dwarf.
{\it Astrophys. J.}, {\bf 524}, 1000$-$1007 (1999).

13.  Stern, D., Bunker, A.~J., Spinrad, H., \& Dey, A.  One-line
redshifts and searches for high-redshift Ly$\alpha$ emission.  {\it
Astrophys.  J.}, {\bf 537}, $73-79$ (2000b).

14.  Stockton, A. \& Ridgway, S.~E.  Deep spectroscopy in the field
of 3C~212.  {\it Astron. J.}, {\bf 115}, $1340-1347$ (1998).

15.  Hogg, D.~W., Cohen, J.~G., Blandford, R., \& Pahre, M.~A.  The
[\ion{O}{2}] luminosity density of the Universe.  {\it Astrophys. J.}, {\bf
504}, $622-628$ (1998).

16.  Leitherer, C., Schaerer, D., Goldader, J.~D., Delgado, R.~M.~G.,
Robert, C., Kune, D.~F., Mello, D.~F.~D., Devost, D., \& Heckman,
T.~M.  Starburst99:  synthesis models for galaxies with active star
formation.  {\it Astrophys. J. Suppl.}, {\bf 123}, $3 - 40$ (1999).

17.  Madau, P.  Radiative transfer in a clumpy universe:  the colors of
high-redshift galaxies.  {\it Astrophys. J.}, {\bf 441}, $18-27$
(1995).

\bigskip

{\noindent \bf Acknowledgments}

\noindent{We gratefully acknowledge Hsiao-Wen Chen and Ken Lanzetta,
who have been generous with information and supportive of our follow-up
efforts on this intriguing source.  We thank Jon Gardner and Josh
Bloom for comments on the STIS data.  We are indebted to the expertise
of the staff of Keck Observatory for their help in obtaining the data
presented herein.  We especially thank Gary Punawai and Jerome for
their assistance during the observing runs.  The work of DS and PE were
carried out at the Jet Propulsion Laboratory, California Institute of
Technology, under a contract with NASA.  The work of WvB and WdV at
IGPP/LLNL was performed under the auspices of the U.S. Department of
Energy by University of California Lawrence Livermore National
Laboratory.  This work has been supported by a grant from the NSF (HS).}

{\noindent Correspondence and requests for materials should be
addressed to D.S.\\ (e-mail:  {\tt
stern@zwolfkinder.jpl.nasa.gov})}.

\eject

% FIGURE 1
\begin{figure}[!t]
\begin{center}
\plotfiddle{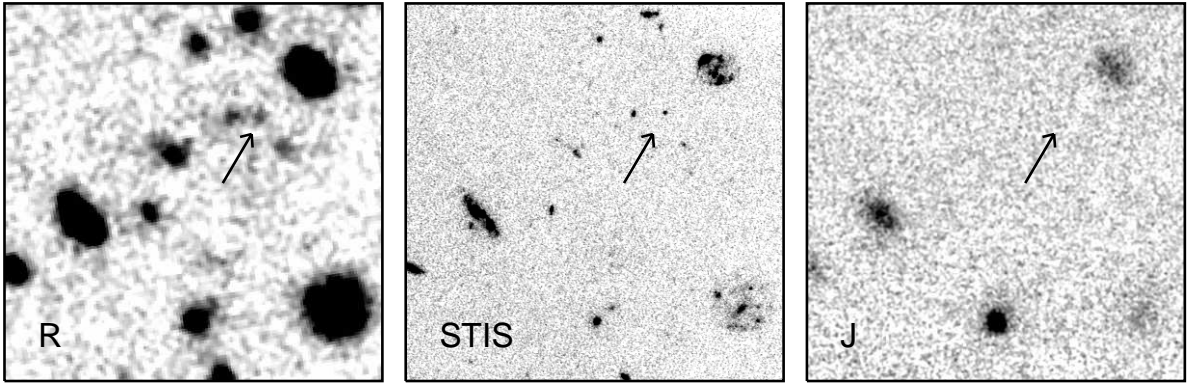}{1.5in}{90}{70}{70}{275}{-55}
\end{center}

\caption{Imaging of \stis\ showing that the galaxy is detected at $R$,
below the Lyman limit if $z = 6.68$.  Fields are 20\farcs0 on a side,
with north at the top and east to the left.  The 70~minute interference
$R$-band ($\lambda_c = 6700$ \AA; $\Delta \lambda = 1400$ \AA) image
was obtained on UT 2000 May 05 with the Echelle Spectrograph and Imager
on the Keck~II telescope in photometric conditions and 0\farcs85
seeing.  The image has been smoothed by a circular Gaussian of width 1
pixel (0\farcs153).  The 64~minute $J$-band ($\lambda_c = 1.25 \mu$m;
$\Delta\lambda = 0.3 \mu$m) image was obtained on UT 2000 January 30
with the Near Infrared Camera$^{7}$ on the Keck~I telescope in
photometric conditions and 0\farcs9 seeing.  A 3\farcs0-long arrow
indicates the location of \stis\ ($\alpha_{\rm J2000} = 12^{\rm h}
36^{\rm m} 27.38^{\rm s}, \delta_{\rm J2000} = +62\deg 17\arcmin
55\farcs56$); it is detected in the $R$-band and filterless (clear)
STIS image, but is undetected in the $J$-band image.  The bright galaxy
NW of \stis\ shows an emission line at 9660 \AA, most likely identified
with either \oii\ at $z = 1.592$ or H$\alpha$ at $z = 0.472$.}

\label{fig_one}
\end{figure}

% FIGURE 2
\begin{figure}[!t]
\begin{center}
\plotfiddle{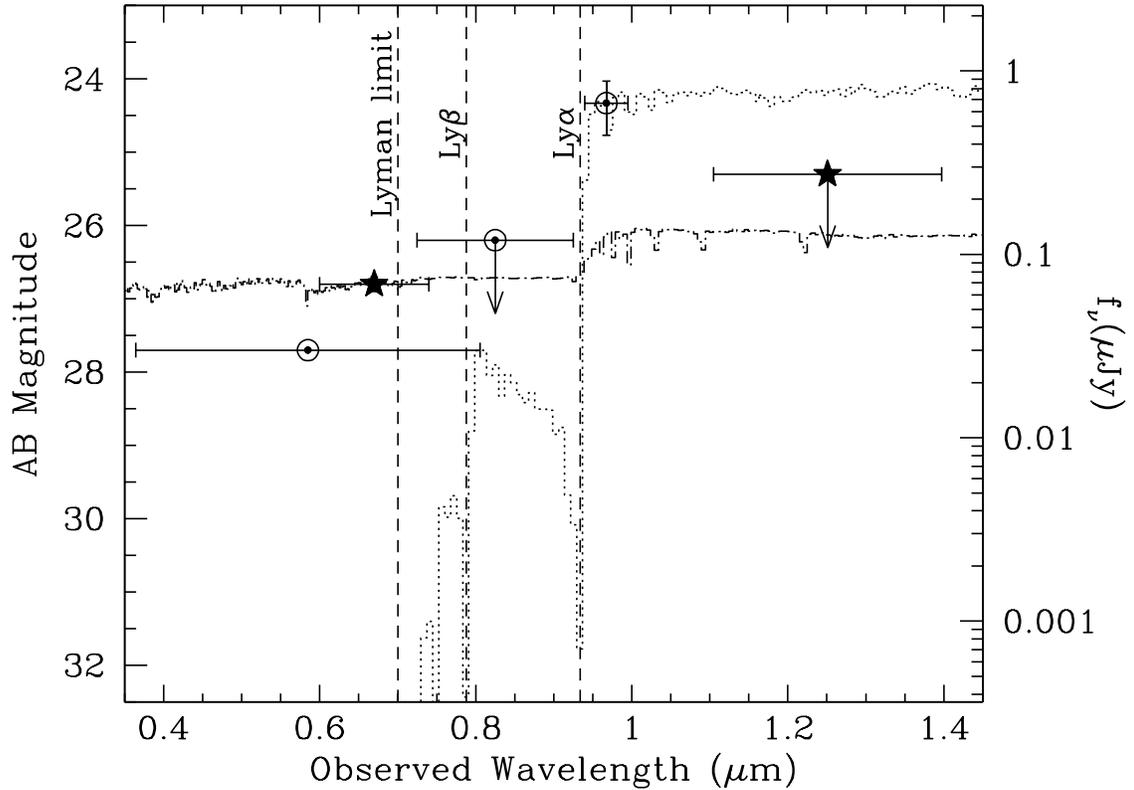}{3.6in}{-90}{60}{60}{-240}{325}
\end{center}

\caption{The photometric detection of \stis\ at 6700\AA\ is not
consistent with $z > 6$.  Solid stars show our $R$- and $J$-band Keck
photometry, where horizontal extent illustrates the width of the
interference filters.  Note that the $R$-band filter has a sharp red
cutoff, unlike conventional $R$-band filters.  Dotted circles are STIS
imaging and spectrophotometric results from Chen \etal$^1$.  Arrows are
3$\sigma$ upper limits.  The dotted and dot-dashed lines refer to 50
Myr old star-forming galaxy models at redshifts of $z = 6.68$ and
$z=1.51$, respectively.  The model$^{16}$ is a 50~Myr old starburst ($Z
= 0.020$, $\alpha_{\rm IMF} = 2.35$, $M_{\rm low} = 1 M_\odot$, $M_{\rm
up} = 100 M_\odot$, nebular lines included) with instantaneous star
formation, attenuated by a model of the intervening intergalactic
medium$^{17}$.  Vertical lines indicate the strong edges due to the
\lya\ forest, the \lyb\ forest, and the Lyman limit (rest-frame 912
\AA) at $z=6.68$.  The Keck photometry does not support the suggested
$z=6.68$ redshift and spectral energy distribution.  In particular, the
detection at 6700 \AA\ requires that \stis\ is not at $z > 6$.  The
ground-based photometric data is more consistent with the emission line
reported at 9337 \AA\ being associated with \oii\ from a star-forming
galaxy at $z = 1.51$.  In this case, the luminosity of \stis\ is
comparable to the Small Magellanic Cloud at a lookback time of $\approx
10$ Gyr.}

\label{fig_two}
\end{figure}

\end{document}